\documentclass[aps,prd,preprint,groupedaddress]{revtex4}
\usepackage[dvipdfmx]{graphicx}
\usepackage{gnuplot-lua-tikz}

\def\tH1{\tilde H_1}
\def\varPhi{\mathit{\Phi}}
\def\Tr{\rm{Tr}}
\def\dag{\dagger}

\def\varLambda{\mathit{\Lambda}}
\def\r2{\sqrt 2}
\def\M#1#2{{\cal M}_{#1#2}^{0(0)}}
\def\Mc#1#2{{\cal M}_{#1#2}^\pm}
\def\O#1i{O_{#1i}}
\def\th#1{\theta_#1}
\def\ta#1{\alpha_#1}
\def\PL{\frac{1-\gamma_5}{2}}
\def\PR{\frac{1+\gamma_5}{2}}
\def\Hi{\tilde H_i^0}
\def\MH#1{\tilde M_{H#1}^{02}}
\def\MHc#1{\tilde M_{H#1}^{\pm 2}}
\def\sigH#1{\sigma(\tilde H_#1^0)}

\def\sigSMH#1{\sigma_{SM}(\tilde H_#1^0)}

\def\PRD#1#2#3{Phys. Rev. {\bf D #1}, #2 (#3)}

\def\PTP#1#2#3{Prog. Theor. Phys. {\bf #1}, #2 (#3)}

\def\EPJC#1#2#3{Eur. Phys. J. {\bf C #1}, #2 (#3)}
\def\PLB#1#2#3{Phys. Lett. {\bf B #1}, #2 (#3)}
\def\PRL#1#2#3{Phys. Rev. Lett. {\bf #1}, #2 (#3)}
\def\PRep#1#2#3{Phys. Rep. {\bf #1}, #2 (#3)}

\def\JHEP#1#2#3{JHEP {\bf #1}, #2 (#3)}
\def\SC#1#2#3{Science {\bf #1}, #2 (#3)}

\begin{document}

\preprint{
OCHA-PP-355
}

\title{
Electric dipole moment of the electron in the model with extra Higgs bosons
}


\author{
Noriyuki Oshimo
}
\affiliation{
Department of Physics,
Ochanomizu University,
Tokyo, 112-8610, Japan
}


\date{\today}

\begin{abstract}

     The Higgs sector which is extended from the standard model 
could generally become an origin of CP violation.   
In the model with two doublet and one triplet Higgs fields on SU(2) symmetry,   
we study the electric dipole moment (EDM) of the electron whose non-vanishing value 
has not yet been established experimentally.  
The Higgs potential is assumed to be consistent with supersymmetric theory, 
though supersymmetric particles are not taken into consideration explicitly.  
While the parameter values of the model are severely constrained by the properties of 
the observed Higgs boson, certain ranges  lead to a magnitude for the electron EDM 
around the experimental upper bound.  
Signatures of the extra Higgs bosons in collision experiments are discussed briefly.

\end{abstract}


\maketitle


\section{Introduction\label{intro} } 

     These years we have been convinced that the standard model (SM) can describe 
also the Higgs sector.   
After discovery of the Higgs boson \cite{exp0},  its properties have been studied 
extensively by experiments \cite{exp}.  
The production and decays proceed as the SM predicts.  
Quantitatively, the cross section via gluon fusion and the widths for the decays 
into $\bar bb$, $WW^*$, $ZZ^*$, and $\gamma\gamma$ are observed to be 
consistent with the SM.   
The mass measures approximately 125 GeV.  

     The Higgs sector, however, may not be exactly the same as the SM.  
Its certain extension could also yield the obtained experimental results.  
It would be necessary to examine the Higgs sector in great detail.  
If the Higgs sector is extended from the SM, some direct or indirect phenomenological 
appearances are expected. 
In particular, the interactions of Higgs bosons with quarks and leptons could violate CP invariance,  
which does not occur in the SM.    
The CP-violating phenomena may become a clue to the extended Higgs sector.  
Although detection of extra Higgs bosons would become unequivocal evidence,   
such an achievement may not be easy as the finding of the SM Higgs boson was not.  

     In this paper, we study the electric dipole moments (EDMs) of the electron and the neutron 
in the Higgs model which consists of one triplet and two doublet fields on SU(2) symmetry.  
Experimentally, non-vanishing value of the EDM has not been observed for the electron $d^e$ \cite{baron} 
or the neutron $d^n$ \cite{pend}, and their magnitudes are bounded from above as 
\begin{eqnarray}
    |d^e| &<& 8.7\times 10^{-29} e{\rm cm},  
\label{bound}    \\ 
    |d^n| &<& 3.0\times 10^{-26} e{\rm cm}.    
\end{eqnarray}
The SM predicts much smaller magnitudes.  
On the other hand, if CP symmetry is not conserved in the Higgs boson interactions,  
the EDMs of the electron and the quarks are generated at two-loop level \cite{barr}.  
The experimental results for the observed Higgs boson impose severe constraints on 
extended Higgs models.  
However, it will be shown that the magnitude of the electron EDM could be as large as 
the present experimental upper bound in sizable parameter regions of our model.  
The neutron EDM can not have such a large magnitude as to be comparable to  
the experiment.  
We also discuss the production and decays of the extra Higgs bosons briefly.  

     For Higgs bosons and their interactions with quarks and leptons, we assume the structure 
which are suggested at the electroweak 
energy scale by the supersymmetric SU(5) grand unification theory.  
However, our discussions are solely performed without taking into consideration 
supersymmetric $R$-odd particles, 
so that the obtained results can also be applied to non-supersymmetric models.
Note that the Higgs sector of the minimal supersymmetric extension of the SM, 
although two doublet fields are contained, respects CP symmetry.    
Its violation occurs only through radiative corrections \cite{pilaftsis}.   
On the other hand, CP invariance is not conserved at tree level 
if the additional triplet field is introduced.  
All the complex coefficients of the Lagrangian cannot be made real by 
redefining particle fields, unless some accidental cancellation is assumed.  
In fact, CP-violating polarization asymmetry could be induced in the two-photon 
decay of the Higgs bosons in our model \cite{oshimo}.  
If supersymmetry is supposed, some non-minimal model would be implied by observation of 
CP violation due to the Higgs sector.  

     In sect. \ref{model} our model is briefly described.  In sect. \ref{edme} 
we obtain the EDMs of the electron and the quarks which are generated by the interactions of the 
Higgs bosons with the $t$ quark and the electron or the quark.  
In sect. \ref{analyses} we perform numerical analyses for the electron EDM.   
Conclusion is given in sect. \ref{conclusion}.  In Appendices some equations and formulae are collected.

\section{Model\label{model}}

     We study the electroweak model which has an extended Higgs sector 
with two doublet and one triplet fields for SU(2).   
It is assumed that the Higgs potential and the Higgs 
interactions with quarks and leptons  
are consistent with supersymmetric theory.  
The supersymmetric $R$-odd particles, however, are not taken into 
consideration explicitly.  
Either their contributions to our discussions can be neglected, or there exists no such particle.  

     The model contains the Higgs fields $H_1$, $H_2$, and $\varPhi$, which transform 
as  $({\bf 2},-1/2)$, $({\bf 2},1/2)$, and $({\bf3}, 0)$ for SU(2)$\times$U(1) gauge symmetry, 
\begin{equation}
   H_1 = \left(
      \matrix{h_1^0 \cr
                   h_1^-}
      \right), \quad    H_2 = \left(
      \matrix{h_2^{+}\cr
                   h_2^0}
      \right),  
\label{doublet}
\end{equation}
\begin{eqnarray}      
    \varPhi &=& \frac{1}{\r2}\left(
     \matrix{ \phi^0     & \r2\phi^+ \cr
                   \r2\phi^-    &    -\phi^0}       
           \right).  
\label{triplet}
\end{eqnarray}
We express the neutral components as   
\begin{equation}      
    h_1^0 = \frac{1}{\r2}(h_R^1+ih_I^1), \quad \quad h_2^0 = \frac{1}{\r2}(h_R^2+ih_I^2),    
\end{equation}
\begin{equation}
      \phi^0 = \frac{1}{\r2}(\phi_R+i\phi_I), 
\end{equation}
where the fields with index $R$ or $I$ are real scalar bosons.  
For convenience, $\tH1$ is defined by 
\[
  \tH1 = \left(
      \matrix{h_1^{- *}\cr
                   -h_1^{0*}}
      \right),
\]
whose transformation property is of $({\bf 2},1/2)$.  
Assuming that electromagnetic symmetry is not broken, we write 
the vacuum expectation values (VEVs) of the neutral Higgs fields as 
\begin{equation}
   \langle h_1^0\rangle = v_1{\rm e}^{i\th1}, \quad 
   \langle h_2^0\rangle = v_2{\rm e}^{i\th2}, \quad 
   \langle\phi^0\rangle = v_0{\rm e}^{i\th0}, 
\label{vev}
\end{equation}
where $v_1$, $v_2$, and $v_0$ denote absolute values.  
The ratio of $v_2$ to $v_1$ is expressed by $\tan\beta=v_2/v_1$.  
These absolute values are constrained by the masses of the $Z$ and $W$ bosons, 
whose relations are given by  
\begin{eqnarray}
 M_Z &=& \frac{1}{\r2}\sqrt{(g^2+g'^2)(v_1^2+v_2^2)}\ , 
 \label{Zmass}  \\
 M_W &=& \frac{1}{\r2}g\sqrt{v_1^2+v_2^2+4v_0^2}\ .  
 \label{Wmass}
\end{eqnarray}
The $\rho$ parameter becomes  
\begin{equation}
 \rho = \frac{g^2+g'^2}{g^2}\frac{M_W^2}{M_Z^2} = \frac{v_1^2+v_2^2+4v_0^2}{v_1^2+v_2^2},  
\label{rho}
\end{equation}
which indicates the relation between the masses of the $Z$ and $W$ bosons.   
With $\tan\beta$ being a free parameter, the values of $v_1$ and $v_2$ are determined.  
We take $v_0$ for 3 GeV, which is almost a maximum value allowed experimentally.  

      The Higgs potential is expressed as 
\begin{equation}
   V=V_0+V_1, 
\label{potential}
\end{equation}
corresponding to the terms at tree level $V_0$ and at one-loop level $V_1$ in supersymmetric theory.  
The first term is given by 
\begin{eqnarray}
   V_0 &=& M_1^2|\tH1|^2 + M_2^2|H_2|^2 + M_3^2\Tr[\varPhi^\dag\varPhi] 
   \nonumber \\
   &+& \left( m_1^2\tH1^\dag H_2 + \frac{1}{2}m_2^2\Tr[\varPhi^2] + {\rm H.c.} \right)                                             
     \nonumber  \\
   &+& \left\{\lambda \mu_H^*(\tH1^\dag\varPhi \tH1 +H_2^\dag\varPhi H_2) 
     + \lambda \mu_\phi^*\tH1^\dag\varPhi^\dag H_2 + m_3\tH1^\dag\varPhi H_2 + {\rm H.c.} \right\} 
    \nonumber \\  
   &+& \frac{1}{8}(g^2+g'^2)\left(|\tH1|^4+|H_2|^4\right) 
   \nonumber \\
   &+& \left\{\frac{1}{4}(g^2-g'^2)+|\lambda|^2\right\}|\tH1|^2|H_2|^2 - \frac{1}{2}(g^2+|\lambda|^2)|\tH1^\dag H_2|^2 
    \nonumber \\
   &+& \frac{1}{2}g^2\Tr[\varPhi^\dag(\varPhi\varPhi^\dag-\varPhi^\dag\varPhi)\varPhi]
    \nonumber  \\
   &+& \left(-\frac{1}{2}g^2+|\lambda|^2\right)\tH1^\dag\varPhi\varPhi^\dag\tH1+\frac{1}{2}g^2\tH1^\dag\varPhi^\dag\varPhi\tH1                                         
    \nonumber \\
    &+& \left(-\frac{1}{2}g^2+|\lambda|^2\right)H_2^\dag\varPhi^\dag\varPhi H_2+\frac{1}{2}g^2H_2^\dag\varPhi\varPhi^\dag H_2,
\label{treepot}
\end{eqnarray}
\[
  M_1^2 = |\mu_H|^2 + {\rm Re}(M_{H1}^2), \quad 
  M_2^2 = |\mu_H|^2 + {\rm Re}(M_{H2}^2), \quad 
  M_3^2 = |\mu_\phi|^2 + {\rm Re}(M_\Phi^2), 
\]
where $g$ and $g'$ stand for the gauge coupling constants for SU(2) and U(1), respectively.  
The parameters $M_{H1}^2$, $M_{H2}^2$, and $M_\Phi^2$ are of mass-squared dimension.  
The dimensionless parameter $\lambda$  and the mass parameters $\mu_H$, $\mu_\varPhi$, and 
$m_i$ ($i$=1-3) all have complex values generally.   
This tree level potential, however, does not describe well the neutral Higgs bosons.  
It is necessary to incorporate additional terms, which may come from radiative 
corrections in supersymmetric theory \cite{okada}.  
In the one-loop potential, the terms which are relevant to the neutral Higgs bosons
are given, in simplified form, by
\begin{eqnarray}
  V_1 &\supset& -\frac{3}{16\pi^2}m_t^4\left(\log\frac{m_t^2}{\varLambda^2}+\frac{1}{2}\right)  
        \nonumber \\
     &+& \frac{3}{32\pi^2}M_{t1}^4\left(\log\frac{M_{t1}^2}{\varLambda^2}+\frac{1}{2}\right) 
     + \frac{3}{32\pi^2}M_{t2}^4\left(\log\frac{M_{t2}^2}{\varLambda^2}+\frac{1}{2}\right).  
 \label{onepot}
\end{eqnarray}
Here, $m_t$ denotes the $t$-quark mass, which is expressed as 
\begin{equation}
m_t ^2= |\eta_t|^2v_2^2, 
\end{equation}
with $\eta_t$ being a coupling constant;   
$M_{ti}^2$ ($i$=1,2) are given by 
\begin{equation}
   M_{t1}^2 = |\eta_t|^2v_2^2 + {\rm Re}(M_Q^2), 
   \quad M_{t2}^2 = |\eta_t|^2v_2^2 + {\rm Re}(M_{U^c}^2), 
 \label{stop}
\end{equation}
where $M_Q^2$ and $M_{U^c}^2$ stand for mass-squared parameters;     
and $\varLambda$ is an appropriate energy scale.  
This energy scale is taken as 
\begin{equation}
       -2m_t^2\left(\log\frac{m_t^2}{\varLambda^2}+1\right)  
     + M_{t1}^2\left(\log\frac{M_{t1}^2}{\varLambda^2}+1\right) 
     + M_{t2}^2\left(\log\frac{M_{t2}^2}{\varLambda^2}+1\right) = 0, 
     \label{scale} 
\end{equation} 
which makes the extremum conditions for the VEVs manageable.  

    In general, the Higgs potential has five complex coefficients $\lambda \mu_H^*$, 
$\lambda \mu_\phi^*$, and $m_i$ ($i$=1-3).  
Although two coefficients can be made real by redefining phases of the fields,  
the others remain complex, leading to CP violation.  
Without loss of generality,  
we can define $m_1^2=|m_1^2|$, $m_2^2=-|m_2^2|$, 
$\lambda \mu_H^*=|\lambda \mu_H^*|{\rm e}^{i\ta1}$, 
$\lambda \mu_\phi^*=|\lambda \mu_\phi^*|{\rm e}^{i\ta2}$, and 
$m_3=|m_3|{\rm e}^{i\ta3}$, taking $m_1^2$ and $m_2^2$ for real.   
Owing to the complex coefficients, the VEVs of the Higgs fields in Eq. (\ref{vev}) 
become complex.  
The extremum conditions for the VEVs are given in Appendix \ref{appa}.  
In our scheme the complex phases $\th1$ and $\th2$ appear as 
a linear combination $\th1+\th2$ $(\equiv \theta)$ in the potential, 
so that only the phase $\theta$ is determined at the vacuum.  

      The mass eigenstates for Higgs bosons are determined by the potential $V$ 
in Eq. (\ref{potential}).  
The mass-squared matrices for the neutral Higgs bosons and 
the charged Higgs bosons are expressed respectively 
by a 6$\times$6 real symmetric matrix ${\cal M}^0$ in Appendix \ref{appb} and 
by a 4$\times$4 Hermitian matrix ${\cal M}^\pm$ in Appendix \ref{appc}.  
These matrices are diagonalized by an orthogonal matrix $O$ and by a unitary matrix $U$, 
\begin{eqnarray}
   O^T{\cal M}^0O &=& {\rm diag}\left(\MH1, \MH2, \MH3, \MH4, \MH5, \MH6\right),  \\
      U^\dag{\cal M}^\pm U &=& {\rm diag}\left(\MHc1, \MHc2, \MHc3, \MHc4 \right), 
\end{eqnarray}
where the eigenvalues are in ascending order.  
The neutral Higgs bosons $\Hi$ and the charged Higgs bosons $\tilde H_i^-$ in mass eigenstates 
are then expressed as 
\begin{eqnarray}
   \Hi &=& O_{1i}h_R^1 + O_{2i}h_R^2 + O_{3i}\phi_R + O_{4i}h_I^1 + O_{5i}h_I^2 + O_{6i}\phi_I,  
                  \\
      \tilde H_i^- &=& U_{1i}^*h_1^- + U_{2i}^*h_2^{+*} + U_{3i}^*\phi^- + U_{4i}^*\phi^{+*}.    
\end{eqnarray}
The Goldstone bosons for spontaneous breaking of SU(2) symmetry are represented by 
$\tilde H_1^0$ and $\tilde H_1^-$, and the values of $\MH1$ and $\MHc1$ vanish.  
In the neutral mass eigenstates, the CP-even and CP-odd fields are mixed.  

     Neglecting the generation mixing , the interaction Lagrangian for the neutral Higgs bosons 
 with the quarks and the charged lepton of the first generation is given by 
\begin{eqnarray}
 \cal L &=& -\frac{m_u}{\r2 v_2}{\overline\psi_u}\left(F_u^i\PL + F_u^{i*}\PR\right)\psi_u\Hi 
      \nonumber \\
   & & -\frac{m_d}{\r2 v_1}{\overline\psi_d}\left(F_d^i\PL + F_d^{i*}\PR\right)\psi_d\Hi 
     \nonumber \\
    & & -\frac{m_e}{\r2 v_1}{\overline\psi_e}\left(F_e^i\PL + F_e^{i*}\PR\right)\psi_e\Hi, 
       \label{interaction}  \\
     F_u^i &=& {\rm e}^{-i\th2}(\O2i + i\O5i),  
     \quad 
       F_d^i = F_e^i = {\rm e}^{-i\th1}(\O1i + i\O4i),  
       \nonumber 
 \end{eqnarray}
where $m_u$, $m_d$, and $m_e$ denote the masses of the $u$ quark, $d$ quark, 
and electron, respectively.  
Even if generation mixing is absent, CP invariance is not respected.  
The interactions for the other generations of quarks and charged leptons are 
obtained by exchanging the masses, 
while the coefficients $F_u^i$ and $F_d^i$ ($F_e^i$) remain the same.    
The interaction Lagrangian for the Higgs bosons and the $W$ or $Z$ boson is given by
\begin{eqnarray}
    {\cal L} &=& gM_W\sqrt{\frac{v_1^2+v_2^2}{v_1^2+v_2^2+4v_0^2}}G_W^iW^{+\mu}W_\mu^-\Hi + 
        \frac{1}{2}\sqrt{g^2+g'^2}M_Z G_Z^i Z^\mu Z_\mu\Hi, 
         \label{int_wz} \\
     G_W^i &=& \cos\beta(\O1i\cos\th1+\O4i\sin\th1) + \sin\beta(\O2i\cos\th2+\O5i\sin\th2) 
           \nonumber \\
                                  & & + \frac{4v_0}{\sqrt{v_1^2+v_2^2}}(\O3i\cos\th0+\O6i\sin\th0),                                
                                  \nonumber \\
      G_Z^i &=& \cos\beta(\O1i\cos\th1 +\O4i\sin\th1) +\sin\beta(\O2i\cos\th2+\O5i\sin\th2).      
      \nonumber       
\end{eqnarray}
The $Z$ boson does not couple to the SU(2)-triplet Higgs fields $\phi_R$ and $\phi_I$.   

\section{Electric dipole moment \label{edme}} 

\begin{figure}
\begin{center}
\includegraphics[width=60mm]{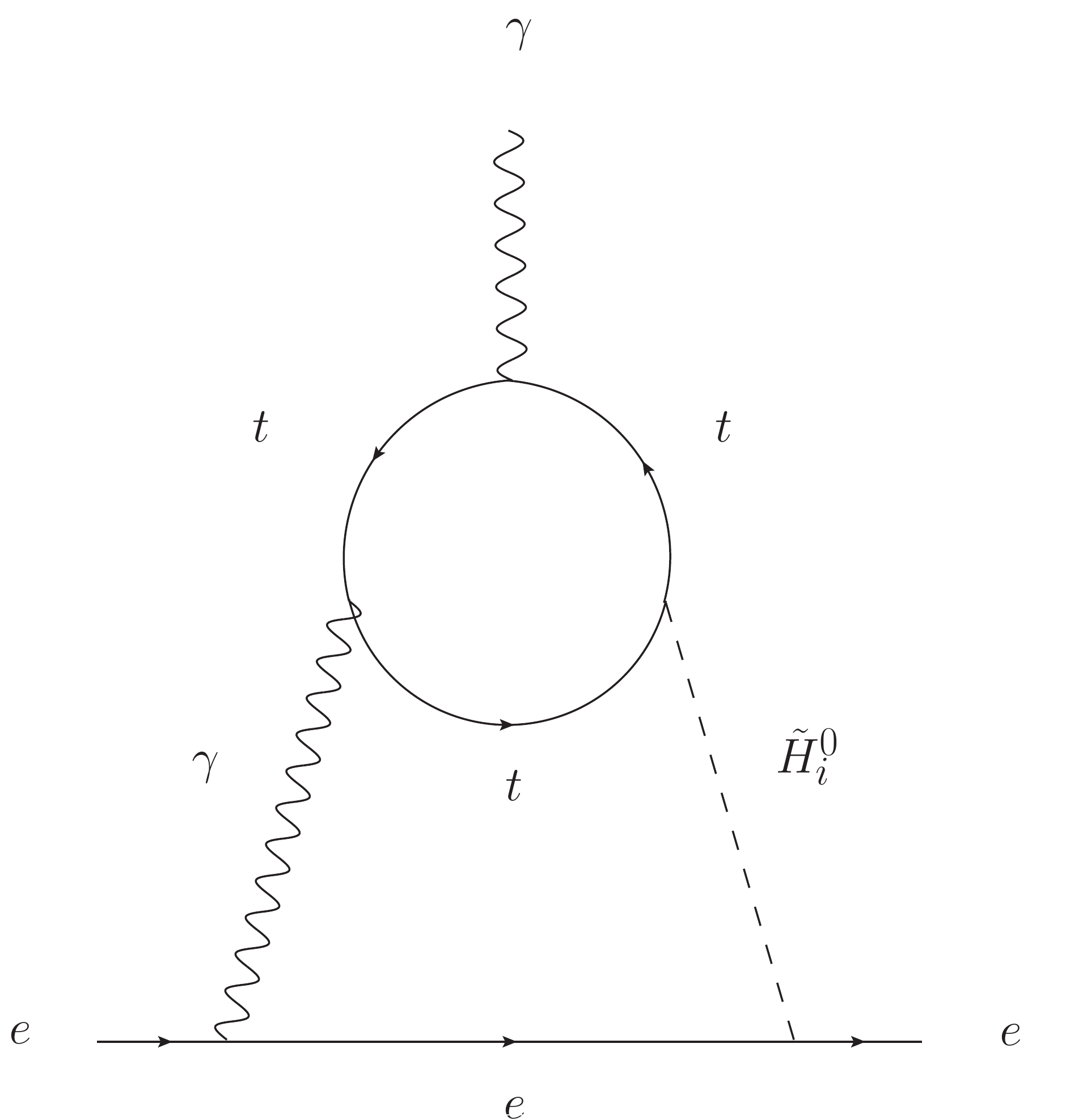}
\end{center}
\caption{The two-loop diagram for the EDM of the electron.  Another diagram in which the photon 
and the Higgs boson lines are interchanged also gives a contribution.}
\label{two-loop}
\end{figure}

     The interactions of the neutral Higgs bosons $\Hi$ ($i$=2-6) and the quarks or 
charged leptons could induce various CP-violating phenomena.  
In particular, sizable effects may be yielded by the processes which quarks of 
the third generation participate in, since the coupling constants are generally 
proportional to the masses and thus non-negligible.  
The EDM is such a quantity. 
The EDM of the electron could be generated at two-loop level by the diagrams mediated 
by the $t$ quark as shown in Fig. \ref{two-loop}. 
Similar diagrams could also lead to non-negligible values for the EDMs of 
the $u$ quark and $d$ quark, 
which may be observed as the EDM of the neutron.  

     The EDM $d^f$ of the electron, $u$ quark, or $d$ quark with electric charge $eQ_f$ 
and mass $m_f$ is given by      
\begin{eqnarray}
   \frac{d^f}{e} &=& \left(\frac{gg'}{12\pi^2}\right)^2\frac{1}{\sin 2\beta}\frac{Q_fm_f}{M_Z^2} 
   \sum_{i=2}^6r_t^i 
   \left[  {\rm Re}(F_f^i){\rm Im}(F_u^i)I(r_t^i) +{\rm Im}(F_f^i){\rm Re}(F_u^i)J(r_t^i,r_f^i)\right], 
  \label{edm} \\
    & & r_t^i=\frac{m_t^2}{\tilde M_{Hi}^{02}}, \quad  r_f^i=\frac{m_f^2}{\tilde M_{Hi}^{02}},  
   \nonumber 
  \end{eqnarray}
where $F_f^i$ denotes an appropriate coupling constant in Eq. (\ref{interaction}), with $e$ being 
the elementary electric charge.  
The functions are defined by 
\begin{eqnarray}
 I(r_t) &=& \frac{1}{2}\int_0^1ds\frac{1-s}{s(1-s)-r_t}\log\frac{s(1-s)}{r_t}, 
 \label{functionI} \\
 J(r_t,r_f) &=& J_1+J_2, 
  \\
  J_1 &=& -\frac{1}{4}\int_0^1ds \frac{1-s}{s(1-s)-r_t}
  \biggl\{s^2 + \biggl[r_t-2+3s-4s^2-\frac{(s-r_t)r_t}{s(1-s)-r_t}\biggr] 
    \log\frac{s(1-s)}{r_t}\biggr\}, 
    \nonumber \\
   J_2 &=& \frac{1}{2}\int\!\!\!\int_Ddsdt\frac{1-s}{\tau^3} r_t \times  \nonumber \\
     & & \Biggl(\sqrt{\frac{\tau}{r_fr_tt}}
   \Bigl\{2+[4r_t-2+s(1-s)]t\Bigr\}\arctan\sqrt{\frac{r_f}{\tau r_tt}}(1-t) 
   \nonumber \\
   & & +\frac{1}{1+\sqrt{\sigma}}\left\{[4(1-r_t)-s(1-s)]t-4-\frac{s(1-s)t}{\sqrt{\sigma}}\right\}\log\frac{1-(1-r_t)t}{r_tt} 
   \nonumber \\
   & & +\frac{1}{\sqrt{\sigma}}\biggl\{\frac{\tau}{r_f}-4+[4(1-r_f)-s(1-s)]t\biggr\} \times  
   \nonumber \\
   & & \quad \quad \quad \quad 
 \left\{ \log\biggl[1-\frac{2r_f(1-t)}{\tau(1+\sqrt{\sigma})}\biggr] - 
         \log\biggl[1+\frac{2r_f(1-t)}{\tau(1+\sqrt{\sigma})}\frac{1-(1-r_t)t}{r_tt}\biggr] \right\}\Biggr), 
   \nonumber \\
  & &  \tau=1-(s^2-s+1)t,   \quad \sigma=1-4r_e\frac{1-(1-r_t)t}{1-(s^2-s+1)t}. 
  \nonumber 
\end{eqnarray}
In these functions the terms which are trivially proportional to $r_f$ have been discarded.  
The integral domain $D$ is approximately given by 
\[
 D:  0\leq s\leq 1, \  0\leq t\leq 1, 
 \]
where small regions proportional to $r_f$ are also neglected.  

     We can evaluate roughly the magnitude of the EDM in Eq. (\ref{edm}).     
The electron mass measures $m_e=5.1\times 10^{-4}$ GeV, so that   
the EDM of the electron is proportional to a factor 
\[
 \left(\frac{gg'}{12\pi^2}\right)^2\frac{1}{\sin 2\beta}\frac{m_e}{M_Z^2} \simeq 4.8\times 10^{-27} {\rm cm}   
 \]
for $\tan\beta=10$.  
This factor is much larger than the experimental bound.  
Therefore, the CP violating phases $\alpha_i$ ($i$=1-3) should have such values as   
to make ${\rm Re}(F_e^i){\rm Im}(F_u^i)$ and ${\rm Im}(F_e^i){\rm Re}(F_u^i)$ 
less than of order of $10^{-2}$.  
Depending on the phases, the predicted magnitude of 
the EDM at two-loop level could have any value below the experimental bound.  

     The EDM of the neutron is described by the EDMs of the quarks.  
The masses of the $u$ quark and $d$ quark measure roughly 
 $m_u=2.2\times 10^{-3}$ GeV and $m_d=4.7\times 10^{-3}$ GeV \cite{pdg}.  
Since the first approximation of the neutron EDM is given by 
$d^n=(4d^d-d^u)/3$, its magnitude is estimated at most to be the experimental bound.  
If the CP violating phases are constrained to keep the EDM of the electron consistent 
with the experiment, the neutron EDM would be much smaller than the bound.  
We therefore do not make numerical analyses for the neutron EDM any more.  

\section{Numerical analyses\label{analyses}}

     The prediction of the EDMs depend on model parameters, which are constrained by 
various experimental results.  
First, by the extremum conditions of the potential in Appendix \ref{appa}, we can express 
the mass-squared parameters ${\rm Re}(M_{H1}^2)$, ${\rm Re}(M_{H2}^2)$, 
${\rm Re}(M_\Phi^2)$, $|m_1^2|$, and $|m_2^2|$ in terms of the VEVs 
of the Higgs bosons $v_1$, $v_2$, $v_0$, $\theta$ $(=\theta_1+\theta_2)$, and 
$\theta_0$ in Eq. (\ref{vev}) and 
the parameters $|\lambda|$, $|\mu_H|$, $|\mu_\phi|$, $|m_3|$, and $\alpha_i$ ($i$=1-3).  
The absolute values $v_1$, $v_2$, and $v_0$ should give the masses of the $W$ and $Z$ bosons 
in Eqs. (\ref{Zmass}) and (\ref{Wmass}).   
The parameters Re($M_Q^2$) and Re($M_{U^c}^2$) in Eq. (\ref{stop}) are left free.  

     A severe constraint comes from the SM Higgs boson which is observed experimentally.  
We take the lightest Higgs boson $\tilde H_2^0$ for this particle.   
Assuming that the Higgs boson is produced dominantly through the gluon fusion 
mediated by the $t$ quark, the cross section $\sigH2$ is proportional 
roughly to the square of the coupling constant for $t$ and $\bar t$.  
The decay widths for $\bar bb$, $WW^*$, and $ZZ^*$ are also proportional to the squares 
of their coupling constants.  
Since the Higgs boson decays dominantly into $b$ and $\bar b$, 
the branching ratios Br($WW^*$) and Br($ZZ^*$) may be given by the 
ratios of their widths to the width for $\bar bb$.  
Therefore, concerning production and decay, the ratios of this model to the SM 
could be estimated roughly by the ratios of the squared coupling constants .  

     The experiments for the Higgs boson have found that its production cross section 
and decay branching ratios are consistent with the prediction by the SM.  
Allowing for uncertainty of our scheme and experimental results, we impose the 
following constraints on the parameters,  
\begin{eqnarray}
  &&\tilde M_{H2}^0 = 120-130 \ \ [\rm GeV], 
  \label{con_mass} \\
  &&\frac{\sigH2}{\sigSMH2} \simeq \frac{v_1^2+v_2^2+4v_0^2}{v_2^2}|F_u^2|^2 = 0.8 - 1.2, 
  \label{tt} \\
  &&\frac{\sigH2\cdot{\rm Br}(WW^*)}{\sigSMH2\cdot{\rm Br}_{SM}(WW^*)} \simeq  
  \cot^2\beta\frac{v_1^2+v_2^2}{v_1^2+v_2^2+4v_0^2}\frac{|F_u^2|^2}{|F_d^2|^2}(G_W^2)^2 
      = 0.8- 1.2,
  \label{ww} \\
   &&\frac{\sigH2\cdot{\rm Br}(ZZ^*)}{\sigSMH2\cdot{\rm Br}_{SM}(ZZ^*)} \simeq 
              \cot^2\beta\frac{|F_u^2|^2}{|F_d^2|^2}(G_Z^2)^ 2= 0.8 - 1.2.  
   \label{zz}
\end{eqnarray}
Here, $\sigma_{SM}$ and ${\rm Br}_{SM}$ denote the cross section and the branching ratio 
under the SM interactions, with $\tilde M_{H2}^0$ being taken for the Higgs boson mass.  
The decay width for $\bar bb$ receives non-negligible contributions from 
QCD corrections \cite{kniehl}.  
However, the above estimate uses the ratio for the two models, which is not affected much 
by the corrections.  
The measurements for the branching ratio of $\tilde H_2^0 \to \gamma\gamma$ give 
values around the SM prediction of $2\times 10^{-3}$.  
The interactions relevant to this decay induce also the production of the Higgs boson and 
its decay into $WW^*$.  
Under the parameter values constrained from Eqs. (\ref{con_mass})-(\ref{zz}), 
the branching ratio Br$(\gamma\gamma)$ becomes compatible with the experimental result.   

\begin{figure}
(a)
\begin{center}
\includegraphics[width=100mm]{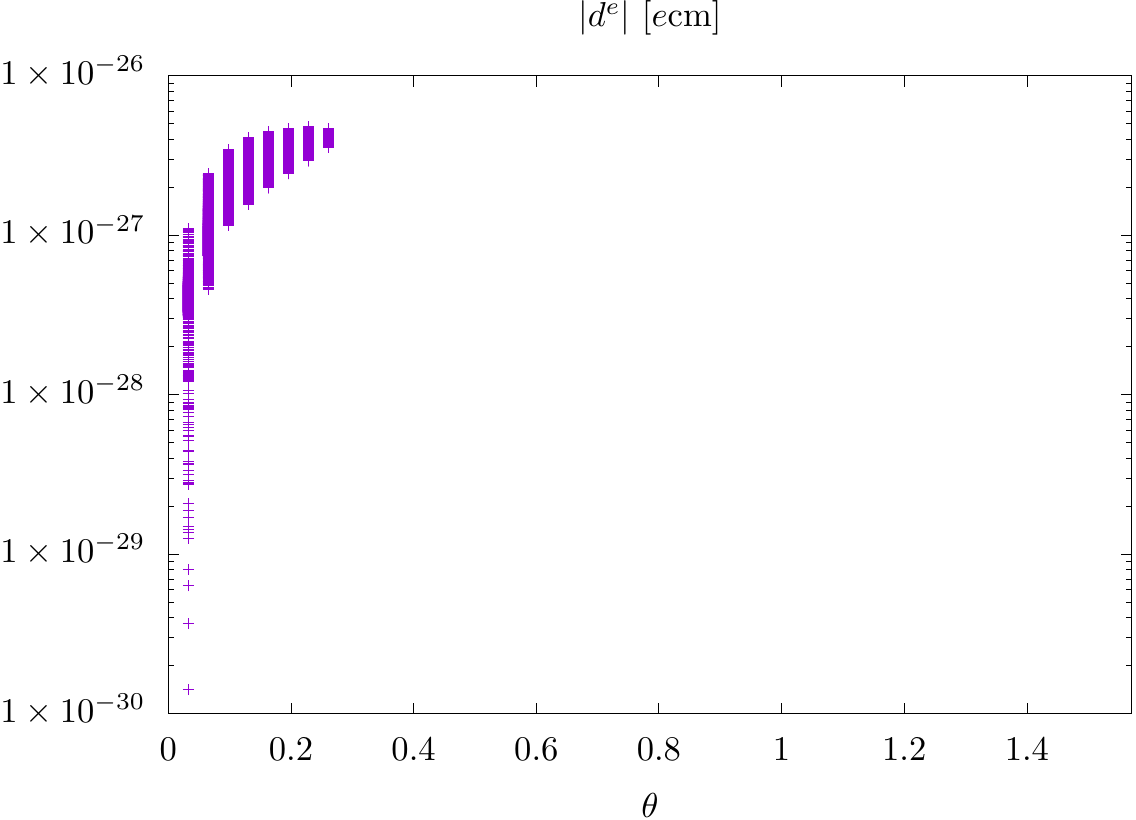}
\end{center}

(b)
\begin{center}
\includegraphics[width=100mm]{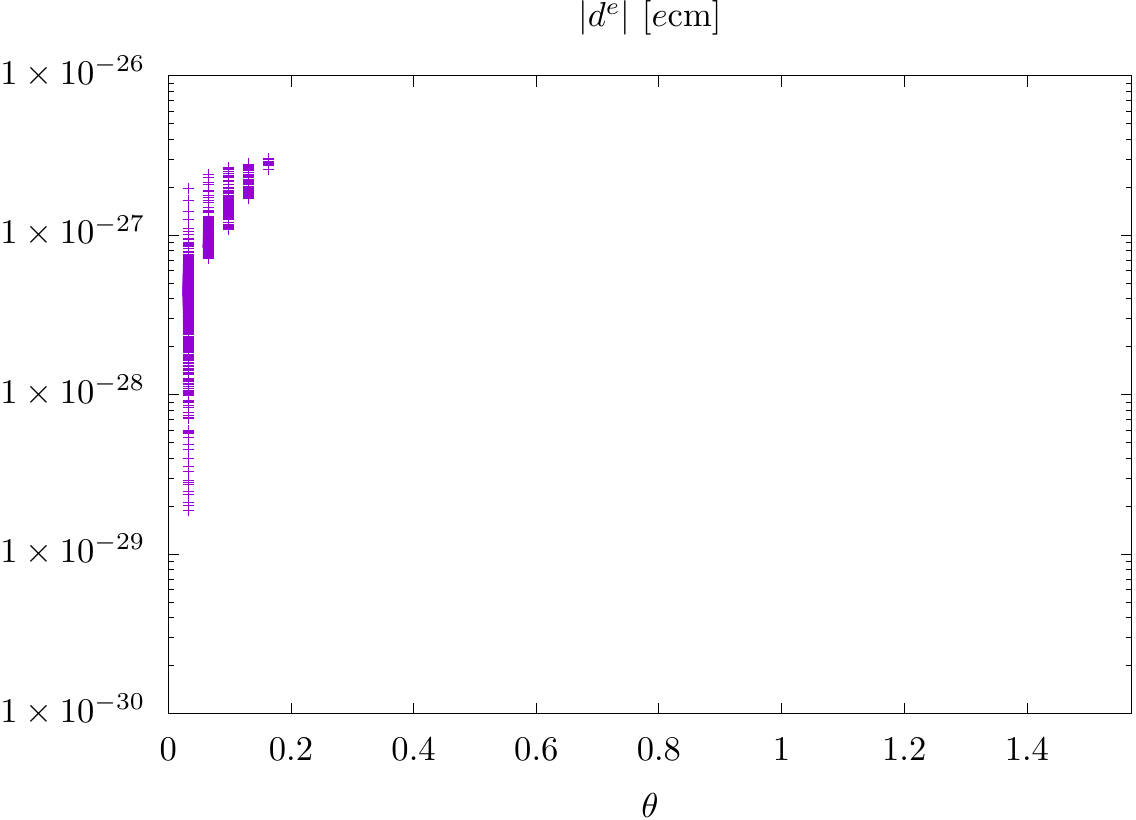}
\caption{The absolute value of the electron EDM as a function of   
the phase $\theta$, with $\theta_0=\pi/12$.  \\
(a) $|\mu_H|=|\mu_\phi|=300$ GeV, (b) $|\mu_H|=|\mu_\phi|=1000$ GeV.
}
\label{figedm}
\end{center}
\end{figure}

      In Fig. \ref{figedm} the absolute value of the electron EDM is shown as a function 
of $\theta$ for $0<\theta<\pi/2$ with $\theta_0=\pi/12$.  
The experimental constraints on the Higgs boson $\tilde H_2^0$ are satisfied 
in sizable regions of the parameter space.  
We show two examples (a) $|\mu_H|=|\mu_\phi|=300$ GeV and (b) $|\mu_H|=|\mu_\phi|=1000$ GeV, 
with $|m_3|={\rm Re}(M_Q^2)={\rm Re}(M_{U^c}^2)=1000$ GeV and $|\lambda|=1$.  
The ratio of the VEVs is fixed at a typical value $\tan\beta=10$.  
With the phases  $\alpha_1$, $\alpha_2$, and $\alpha_3$ 
being varied from $0$ to $2\pi$, if the vacuum is consistent with the experimental 
constraints, excepting the upper bound on the EDM  in Eq. (\ref{bound}), 
the absolute value of the EDM is depicted as a point.  
As the complex phases of the VEVs increase, the magnitude of the EDM becomes large.   
The predicted magnitude is inside the experimental bound for smaller values of $\theta$, 
while outside for larger values.  
For $0.3<\theta$ and $0.2<\theta$ in Figs. \ref{figedm} (a) and \ref{figedm} (b), respectively, 
the parameter values do not satisfy the experimental results for the observed Higgs boson.  

     In Table \ref{phases} two examples are shown for specific phase values.  
The other parameter values of examples (a) and (b) are the same as (a) and (b) 
of Fig. \ref{figedm}, respectively.   
In general, the parameter values of the model is constrained severely by the experimental results 
for the observed Higgs boson.  In spite of this constraint, within the allowed region, 
the EDM can have a magnitude around the present experimental upper bound.  

\begin{table}
\caption{
The values of CP-violating phases and the electron EDM.  
(a) $|\mu_H|=|\mu_\phi|=300$ GeV, (b) $|\mu_H|=|\mu_\phi|=1000$ GeV.  
\label{phases}
}
\vspace{1 cm}
\begin{ruledtabular}
\begin{tabular}{ccccccc}
   & $\alpha_1$ & $\alpha_2$ & $\alpha_3$ & $\theta_0$ & $\theta$ & $|d^e|$ [$e$cm] \\ 
\hline
(a) & -$\frac{3}{8}\pi$ & $\frac{5}{6}\pi$ & -$\frac{1}{8}\pi$ & $\frac{1}{12}\pi$ & $\frac{1}{96}\pi$ 
                                                  & 8.4$\times 10^{-29}$ \\
(b) & -$\frac{1}{2}\pi$ & $\frac{11}{48}\pi$ & -$\frac{47}{48}\pi$ & $\frac{1}{12}\pi$ & $\frac{1}{96}\pi$ 
                                                  & 5.9$\times 10^{-29}$ \\
\end{tabular}
\end{ruledtabular}
\end{table}

     In the present model there exist four extra neutral Higgs bosons $\Hi$ ($i$=3-6).   
The cross sections are roughly proportional to the squares of the coupling coefficients 
with the $t$ quark or with the $b$ quark in Eq. (\ref{interaction}).  
The ratios of these squares to those corresponding to the SM are given by 
\begin{equation}
R_t=\frac{v_1^2+v_2^2+4v_0^2}{v_2^2}|F_u^i|^2, \quad 
R_b=\frac{v_1^2+v_2^2+4v_0^2}{v_1^2}|F_d^i|^2,  
\end{equation}
for the $t$ and $b$ quarks.  
In Table \ref{ratios} these ratios and the branching ratios for $\Hi\to\bar bb, \bar tt, W^+W^-, ZZ$ 
are given, together with the mass values, for the examples (a) and (b).  
Compared to the SM interactions, the magnitude of the interaction 
with the $t$ quark is small, while that with the $b$ quark is large.  
Therefore, the cross sections of the gluon fusion $gg\to \Hi$ mediated by 
the $t$ quark are suppressed, while the cross sections of the gluon fusion mediated 
by the $b$ quark and of the associated production $gg\to \bar bb\Hi$ are enhanced.  
A naive estimate gives that the cross section would be small by a factor of order of $10^{-1}$ 
compared to the $t$-quark mediated gluon fusion in the SM.  
The decay properties are also different from the SM.  
The decay into $\bar bb$ has the largest branching ratio, while in the SM the Higgs-like boson 
with a mass larger than about 200 GeV decays dominantly into $W^+W^-$ and $ZZ$.  

     Experiments have not observed a Higgs-like boson, except the observed one, 
for the mass range smaller than 1000 GeV \cite{higgs3}, provided that 
the phenomena are described like the Higgs boson of the SM.   
Although some of the extra Higgs bosons have mass values within the excluded range, 
the production cross sections become generally small and 
the branching ratios are different from the SM.  
The experimental negative results could not apply to the extra bosons.
     
     This model predicts also three charged Higgs bosons.  
Experiments have not found such a boson, and the lower bound 
on the mass is obtained as 80 GeV \cite{chiggs}.  
On the other hand, the masses of the charged Higgs bosons in our model become generally 
larger than 100 GeV.  
In Table \ref{cmasses} the mass values are listed for examples (a) and (b).  
There may exist a charged Higgs boson whose mass is not much 
above the excluded range.      

\begin{table}
\caption{
The masses, branching ratios, and ratios for coupling strengths of the 
neutral Higgs bosons for examples (a) and (b).  The lightest particle corresponds to 
the observed Higgs boson.  
\label{ratios}
}
\vspace{0.5cm}
\begin{ruledtabular}
\begin{tabular}{ccccccc}
   mass [GeV] & $\bar bb$ & $\bar tt$  & $W^+W^-$ & $ZZ$ & $R_t$ & $R_b$ \\
\hline
(a) &&&&&& \\
\hline
 123 & 1.0                           & 0                                    & 0                              & 0 
                                                                                     & 9.9$\times 10^{-1}$          & 1.0 \\
 178 & 9.7$\times 10^{-1}$ & 0                              & 3.4$\times 10^{-2}$ & 0 
                                                                       & 1.5$\times 10^{-2}$ & 9.4$\times 10$  \\
 222 & 9.9$\times 10^{-1}$ & 0                              & 4.5$\times 10^{-3}$ & 1.8$\times 10^{-3}$ 
                                                                       & 9.3$\times 10^{-3}$ & 1.0$\times 10^{2}$ \\
 607 & 4.4$\times 10^{-1}$ & 2.0$\times 10^{-1}$ & 2.6$\times 10^{-1}$ & 9.7$\times 10^{-2}$ 
                                                                       & 3.1$\times 10^{-3}$ & 6.4  \\
 247$\times 10$ & 8.9$\times 10^{-3}$ & 2.4$\times 10^{-2}$ & 6.5$\times 10^{-1}$ & 3.2$\times 10^{-1}$ 
                                                                              & 1.1$\times 10^{-4}$ & 6.5$\times 10^{-2}$ \\
 \hline
 (b) &&&&&& \\
\hline
 121 & 1.0                           & 0                                    & 0                              & 0 
                                                                                     & 1.0                           & 1.0 \\
 193 & 1.0                           & 0                              & 5.1$\times 10^{-8}$ & 4.8$\times 10^{-4}$ 
                                                                       & 1.1$\times 10^{-2}$ & 7.9$\times 10$  \\
 281 & 1.0                           & 0                              & 1.6$\times 10^{-4}$ & 7.0$\times 10^{-5}$ 
                                                                       & 9.0$\times 10^{-3}$ & 1.0$\times 10^{2}$ \\
 478 & 8.2$\times 10^{-1}$ & 8.3$\times 10^{-2}$ & 6.9$\times 10^{-2}$ & 2.9$\times 10^{-2}$ 
                                                                       & 1.9$\times 10^{-3}$ & 2.1$\times 10$  \\
 497$\times 10$ & 2.8$\times 10^{-4}$ & 5.3$\times 10^{-3}$ & 6.6$\times 10^{-1}$ & 3.3$\times 10^{-1}$ 
                                                                              & 9.0$\times 10^{-5}$ & 8.1$\times 10^{-3}$ \\
\end{tabular}
\end{ruledtabular}
\
\end{table}

\begin{table}
\caption{
The masses [GeV] of the charged Higgs bosons for examples (a) and (b).  
\label{cmasses}
}
\vspace{1 cm}
\begin{ruledtabular}
\begin{tabular}{cccc}
(a) & 222 & 608 & 247$\times 10$\\
(b) & 235 & 483 & 497$\times 10$ \\
\end{tabular}
\end{ruledtabular}
\end{table}

\section{Conclusion\label{conclusion}}
 
      We have studied the EDM of the electron, assuming the extension of the SM 
which has the Higgs fields of two doublet and one triplet representations 
for SU(2) transformation.  
The extended Higgs sector induces naturally violation of CP invariance at tree level.  
As a result, charged leptons and quarks could have non-vanishing EDMs at two-loop level, 
which are mediated by the Higgs bosons and the $t$ quark.  
On the other hand, the extended sector is severely constrained from 
the experimental results for the observed Higgs boson.   
The EDM of the electron has not been observed and its upper bound 
on the magnitude is obtained.  
We have found the region of parameter space 
which is consistent with the Higgs boson and gives a large magnitude for the EDM.  
The EDM can be expected to have a magnitude around 
the present experimental bound.  

     Possible CP violating effects in the SM are very restricted.  
If the Higgs sector does not conserve CP invariance, the resultant phenomena would 
be easily distinguished from the SM.  
One example is the EDM of the electron or the neutron.  
The two-photon decay of the Higgs boson may also show CP asymmetry for the helicities.  
Important clues for physics beyond the SM may be provided by 
examining CP violation.  

     Our model predicts extra neutral Higgs bosons, which could have escaped detection 
in experiments.  
The production and decay properties of these bosons are 
different much from the SM Higgs boson.  
Their production cross sections by the gluon fusion generally become smaller.  
The dominant decay modes are $\bar bb$, and not $W^+W^-$ and $ZZ$,  
even if kinematically allowed.   
In order to detect the extra Higgs bosons, it would be necessary to make experimental 
analyses which are different from those for searching a Higgs-like boson of the SM.

\appendix
\section{Extremum conditions \label{appa}} 

     The extremum conditions $\partial V/\partial v_1$, $\partial V/\partial v_2$, $\partial V/\partial v_0$, 
$\partial V/\partial\theta$, and $\partial V/\partial\theta_0$ are given by 
\begin{eqnarray}
   && M_1^2 + \frac{g^2+g'^2}{4}v_1^2 -\Bigl(\frac{g^2+g'^2}{4}-\frac{|\lambda|^2}{2}\Bigr)v_2^2 
          + \frac{|\lambda|^2}{2}v_0^2 - 
                                         \r2 v_0|\lambda \mu_H^*|\cos(\ta1+\th0) = 
      \nonumber \\
 &&   \frac{v_2}{v_1}\left\{|m_1^2|\cos\theta  - \frac{v_0}{\r2}\left[|\lambda \mu_\phi^*|\cos(\ta2-\th0+\theta) 
                                              + |m_3|\cos(\ta3+\th0+\theta)\right]\right\}, 
       \label{extremum1}\\
   && M_2^2 + \frac{g^2+g'^2}{4}v_2^2 -\Bigl(\frac{g^2+g'^2}{4}-\frac{|\lambda|^2}{2}\Bigr)v_1^2 
                + \frac{|\lambda|^2}{2}v_0^2 - 
                                        \r2 v_0|\lambda\mu_H^*|\cos(\ta1+\th0) = 
      \nonumber \\
&&   \frac{v_1}{v_2}\left\{|m_1^2|\cos\theta  - \frac{v_0}{\r2}\left[|\lambda \mu_\phi^*| \cos(\ta2-\th0+\theta)
                                                + |m_3| \cos(\ta3+\th0+\theta)\right]\right\}, 
 \label{extremum2} \\
  && M_3^2 + \frac{|\lambda|^2}{2}(v_1^2+v_2^2) -|m_2^2|\cos2\th0 
                                              - \frac{v_1^2+v_2^2}{\r2 v_0}|\lambda\mu_H^*|\cos(\ta1+\th0) =  
         \nonumber \\
  &&    \ \  -\frac{v_1v_2}{\r2 v_0}\left\{|\lambda m_\phi^*|\cos(\ta2-\th0+\theta) + 
                  |m_3|\cos(\ta3+\th0+\theta)\right\},
  \label{extremum3}\\
  && |m_1^2|\sin\theta  = \frac{v_0}{\r2}\left\{|\lambda \mu_\phi^*|\sin(\ta2-\th0+\theta)
                                                                              + |m_3|\sin(\ta3+\th0+\theta)\right\},
 \label{extremum4}\\
  && |m_2^2|\sin2\th0  + \frac{v_1^2+v_2^2}{\r2 v_0}|\lambda \mu_H^*|\sin(\ta1+\th0) =   
         \nonumber \\
  &&    \ \  -\frac{v_1v_2}{\r2 v_0}\left\{|\lambda \mu_\phi^*|\sin(\ta2-\th0+\theta) 
                                    - |m_3|\sin(\ta3+\th0+\theta)\right\}.       
          \label{extremum5}                    
\end{eqnarray}
Since the equations 
$\langle\partial V_1/\partial h_R^2\rangle =  \langle\partial V_1/\partial h_I^2\rangle = 0$ 
are derived from the assumption in Eq.~(\ref{scale}), 
these extremum conditions are the same as those for the 
tree-level potential $V_0$.  

\section{Mass-squared matrix for the neutral Higgs bosons \label{appb}} 

    The mass-squared matrix ${\cal M}^0$ for the neutral Higgs bosons 
receives contributions from the tree-level potential and the one-loop potential, 
${\cal M}^0={\cal M}^{0(0)}+{\cal M}^{0(1)}$.  
The elements $M_{ij}^0$ are given by 
\begin{eqnarray}
\M11 &=& M_1^2 + \frac{g^2+g'^2}{4}(1+2\cos^2\th1)v_1^2 
              -\Bigl(\frac{g^2+g'^2}{4}-\frac{|\lambda|^2}{2}\Bigr)v_2^2           \nonumber \\
                & & +\frac{|\lambda|^2}{2}v_0^2 - \r2 v_0|\lambda\mu_H^*|\cos(\ta1+\th0), 
    \\
\M12 &=& -\Bigl(\frac{g^2+g'^2}{2}-|\lambda|^2\Bigr)v_1v_2\cos\th1\cos\th2 - |m_1^2 | 
      \nonumber   \\
         & & +\frac{v_0}{\r2}\{|\lambda\mu_\phi^*| \cos(\ta2-\th0) + |m_3|\cos(\ta3+\th0)\},
   \\
\M13 &=& |\lambda|^2v_1v_0\cos\th1\cos\th0- \r2 v_1 |\lambda\mu_H^*|\cos\ta1\cos\th1 
    \nonumber   \\
        & & +\frac{v_2}{\r2}\{|\lambda\mu_\phi^*|\cos(\ta2+\th2) + |m_3|\cos(\ta3+\th2)\},
   \\
\M14 &=& \frac{g^2+g'^2}{4}v_1^2\sin2\th1, 
 \\
\M15 &=& -\Bigl(\frac{g^2+g'^2}{2}-|\lambda|^2\Bigr)v_1v_2\cos\th1\sin\th2 
      \nonumber \\
         & & - \frac{v_0}{\r2}\{|\lambda\mu_\phi^*|\sin(\ta2-\th0) + |m_3|\sin(\ta3+\th0)\}, 
  \\
\M16 &=& |\lambda|^2v_1v_0\cos\th1\sin\th0 + \r2 v_1|\lambda\mu_H^*|\sin\ta1\cos\th1  
    \nonumber   \\
       & & + \frac{v_2}{\r2}\{|\lambda\mu_\phi^*|\sin(\ta2+\th2) - |m_3|\sin(\ta3+\th2)\}, 
  \\
\M22 &=& M_2^2 + \frac{g^2+g'^2}{4}v_2^2(1+2\cos^2\th2) 
               -\Bigl(\frac{g^2+g'^2}{4}-\frac{|\lambda|^2}{2}\Bigr)v_1^2         \nonumber \\
            & &   + \frac{|\lambda|^2}{2}v_0^2 - \r2v_0|\lambda\mu_H^*|\cos(\ta1+\th0),
       \\
\M23 &=& |\lambda|^2v_2v_0\cos\th2\cos\th0 - \r2 v_2|\lambda\mu_H^*|\cos\ta1\cos\th2
    \nonumber   \\
       & & + \frac{v_1}{\r2}\{|\lambda\mu_\phi^*|\cos(\ta2+\th1) + |m_3|\cos(\ta3+\th1)\}, 
  \\
\M24 &=& -\Bigl(\frac{g^2+g'^2}{2}-|\lambda|^2\Bigr)v_1v_2\sin\th1\cos\th2 
    \nonumber   \\
       & & - \frac{v_0}{\r2}\{|\lambda\mu_\phi^*|\sin(\ta2-\th0) + |m_3|\sin(\ta3+\th0)\}, 
  \\
\M25 &=& \frac{g^2+g'^2}{4}v_2^2\sin2\th2, 
 \\
\M26 &=& |\lambda|^2v_2v_0\cos\th2\sin\th0 + \r2 v_2|\lambda\mu_H^*|\sin\ta1\cos\th2 
    \nonumber   \\
       & & + \frac{v_1}{\r2}\{|\lambda\mu_\phi^*|\sin(\ta2+\th1) - |m_3|\sin(\ta3+\th1)\}, 
  \\
\M33 &=& M_3^2 + \frac{|\lambda|^2}{2}(v_1^2+v_2^2) - |m_2^2|
      \\
\M34 &=& |\lambda|^2v_1v_0\sin\th1\cos\th0 - \r2 v_1|\lambda\mu_H^*|\cos\ta1\sin\th1  
   \nonumber    \\
       & & - \frac{v_2}{\r2}\{|\lambda\mu_\phi^*|\sin(\ta2+\th2) + |m_3|\sin(\ta3+\th2)\}, 
  \\
\M35 &=& |\lambda|^2v_2v_0\sin\th2\cos\th0 - \r2 v_2|\lambda\mu_H^*|\cos\ta1\sin\th2 
    \nonumber   \\
       & & - \frac{v_1}{\r2}\{|\lambda\mu_\phi^*|\sin(\ta2+\th1) + |m_3|\sin(\ta3+\th1)\}, 
  \\
\M36 &=& 0,
  \\
\M44 &=& M_1^2 + \frac{g^2+g'^2}{4}v_1^2(1+2\sin^2\th1) 
           -\Bigl(\frac{g^2+g'^2}{4}-\frac{|\lambda|^2}{2}\Bigr)v_2^2              \nonumber \\
                     & &  +\frac{|\lambda|^2}{2}v_0^2 - \r2v_0|\lambda\mu_H^*|\cos(\ta1+\th0),
       \\
\M45 &=& -\Bigl(\frac{g^2+g'^2}{2}-|\lambda|^2\Bigr)v_1v_2\sin\th1\sin\th2 + |m_1^2|  
   \nonumber   \\
         & & -\frac{v_0}{\r2}\{|\lambda\mu_\phi^*|\cos(\ta2-\th0) + |m_3|\cos(\ta3+\th0)\},
   \\
\M46 &=& |\lambda|^2v_1v_0\sin\th1\sin\th0 + \r2 v_1|\lambda\mu_H^*|\sin\ta1\sin\th1  
    \nonumber   \\
       & & + \frac{v_2}{\r2}\{|\lambda\mu_\phi^*|\cos(\ta2+\th2) - |m_3|\cos(\ta3+\th2)\}, 
  \\
\M55 &=& M_2^2 + \frac{g^2+g'^2}{4}v_2^2(1+2\sin^2\th2) 
               -\Bigl(\frac{g^2+g'^2}{4}-\frac{|\lambda|^2}{2}\Bigr)v_1^2            \nonumber \\                         
                   & &   +\frac{|\lambda|^2}{2}v_0^2 - \r2 v_0|\lambda\mu_H^*|\cos(\ta1+\th0),
       \\
\M56 &=& |\lambda|^2v_2v_0\sin\th2\sin\th0 + \r2 v_2|\lambda\mu_H^*|\sin\ta1\sin\th2  
   \nonumber    \\
       & & + \frac{v_1}{\r2}\{|\lambda\mu_\phi^*|\cos(\ta2+\th1) - |m_3|\cos(\ta3+\th1)\}, 
  \\
\M66 &=& M_3^2 + \frac{|\lambda|^2}{2}(v_1^2+v_2^2) + |m_2^2|, 
  \\
{\cal M}_{22}^{0(1)} &=& \frac{3}{8\pi^2}\frac{m_t^4}{v_2^2}\cos^2\th2\log\frac{M_{t1}^2M_{t2}^2}{m_t^4}, 
\\
{\cal M}_{55}^{0(1)} &=& \frac{3}{8\pi^2}\frac{m_t^4}{v_2^2}\sin^2\th2\log\frac{M_{t1}^2M_{t2}^2}{m_t^4}, 
\\
{\cal M}_{25}^{0(1)} &=& 
                  \frac{3}{8\pi^2}\frac{m_t^4}{v_2^2}\sin\th2\cos\th2\log\frac{M_{t1}^2M_{t2}^2}{m_t^4}, 
\end{eqnarray}
where the extremum conditions in Eqs. (\ref{extremum1})-(\ref{extremum5}) are not taken into account.  
The indices $i, j$ (=1-6) are in order of $(h_R^1,h_R^2, \phi_R, h_I^1, h_I^2, \phi_I)$.

\section{Mass-squared matrix for the charged Higgs bosons \label{appc}} 

    The mass-squared matrix $\cal M^\pm$ for the charged Higgs bosons receives contributions dominantly from 
the tree-level potential.  The elements $M_{ij}^\pm$ are given by 
\begin{eqnarray}
\Mc11 &=& M_1^2 + \frac{g^2+g'^2}{4}v_1^2 + \Bigl(\frac{g^2-g'^2}{4}+|\lambda|^2\Bigr)v_2^2 
             + \frac{|\lambda|^2}{2}v_0^2              \nonumber \\
             & & + \r2|\lambda\mu_H^*|v_0\cos(\ta1+\th0), 
    \\
\Mc12 &=& \frac{g^2+|\lambda|^2}{2}v_1v_2\exp[i(\th1+\th2)]+ |m_1^2 | 
      \nonumber   \\
          & & +\frac{v_0}{\r2}\{|\lambda\mu_\phi^*| \exp[-i(\ta2-\th0)] + |m_3|\exp[-i(\ta3+\th0)]\},
   \\
\Mc13 &=& \frac{g^2-|\lambda|^2}{\r2}v_1v_0\exp[i(\th1 -\th0)]
       - v_1 |\lambda\mu_H^*|\exp[i(\ta1 +\th1)]            \nonumber \\
       & &  + v_2|\lambda\mu_\phi^*|\exp[-i(\ta2+\th2)],
   \\
\Mc14 &=& -\frac{g^2-|\lambda|^2}{\r2}v_1v_0\exp[i(\th1 +\th0)]
       - v_1 |\lambda\mu_H^*|\exp[-i(\ta1 -\th1)]         \nonumber \\
         & & + v_2|m_3|\exp[-i(\ta3+\th2)],
   \\
\Mc22 &=& M_2^2 + \Bigl(\frac{g^2-g'^2}{4}+|\lambda|^2\Bigr)v_1^2 + \frac{g^2+g'^2}{4}v_2^2 
                    + \frac{|\lambda|^2}{2}v_0^2        \nonumber \\
                  & &  +  \r2v_0|\lambda\mu_H^*|\cos(\ta1+\th0),
       \\
\Mc23 &=& \frac{g^2-|\lambda|^2}{\r2}v_2v_0\exp[-i(\th2 +\th0)]
       + v_2 |\lambda\mu_H^*|\exp[i(\ta1 -\th2)]       \nonumber \\
        & &  - v_1|m_3|\exp[i(\ta3+\th1)],
   \\
\Mc24 &=& -\frac{g^2-|\lambda|^2}{\r2}v_2v_0\exp[-i(\th2 -\th0)]
       + v_2 |\lambda\mu_H^*|\exp[-i(\ta1 +\th2)]       \nonumber \\
        & &   - v_1|\lambda\mu_\phi^*|\exp[i(\ta2+\th1)],
   \\
\Mc33 &=& M_3^2 - (\frac{g^2}{2}-|\lambda|^2)v_1^2+\frac{g^2}{2}v_2^2 + g^2v_0^2,
      \\
\Mc34 &=& -g^2v_0^2\exp[2i\th0]-|m_2^2|, \\
\Mc44 &=& M_3^2 +\frac{g^2}{2}v_1^2 - (\frac{g^2}{2}-|\lambda|^2)v_2^2+  g^2v_0^2,
\end{eqnarray}
where the extremum conditions in Eqs. (\ref{extremum1})-(\ref{extremum5}) are not taken into account.  
The indices $i, j$ (=1-4) are in order of $(h_1^-,h_2^{+*}, \phi^-, \phi^{+*})$.


\end{document}